\documentclass[fleqn,usenatbib]{mnras}

\usepackage{newtxtext,newtxmath}
\usepackage[T1]{fontenc}
\usepackage{graphicx}
\usepackage{amsmath}
\usepackage{dcolumn}
\usepackage{bm}
\usepackage[multiple]{footmisc}
\usepackage{multirow}

\title{A phenomenological model for dark matter phase space distribution}

\author[L. Zhen \& H. Steen]{
Zhen Li, $^{1}$\thanks{zhen.li@nbi.ku.dk; CA }
Steen H. Hansen, $^{1}$\thanks{hansen@nbi.ku.dk}
\\
$^{1}$ DARK, Niels Bohr Institute, University of Copenhagen, Jagtvej 128, 2200 Copenhagen Ø, Denmark\\
}

\date{Accepted XXX. Received YYY; in original form ZZZ}

\pubyear{2024}

\begin{document}
\label{firstpage}
\pagerange{\pageref{firstpage}--\pageref{lastpage}}
\maketitle

\begin{abstract}
Understanding the nature of dark matter is among the top priorities of modern physics. However, due to its inertness, detecting and studying it directly in terrestrial experiments is extremely challenging. Numerical N-body simulations currently represent the best approach for studying the particle properties and phase space distribution, assuming the collisionless nature of dark matter. These simulations also address the lack of a satisfactory theory for predicting the universal properties of dark matter halos, including the density profile and velocity distribution. In this work, we propose a new phenomenological model for the dark matter phase space distribution. This model aims to provide an NFW-like density profile, velocity magnitude distribution, and velocity component distributions that align closely with simulation data.  Our model is relevant both for theoretical modeling of dark matter distributions, as well as for underground detector experiments that rely on the dark matter velocity distribution for experimental analysis.
\end{abstract}


\section{Introduction}\label{sec1}

Dark matter is an invisible and mysterious substance that makes up a great portion of the universe. While its existence is solely inferred from its gravitational effects, its true nature and composition remain one of the most significant questions in modern astrophysics and particle physics. Detecting dark matter has proven to be incredibly challenging because it does not emit or interact with light. Based on these limitations, cosmological N-body simulations may, at the moment, be the only possibly way to study dark matter structures and distributions. These simulations are performed assuming the collisionless property of dark matter, see the review of simulations \cite{rev,rev1}.

Across a wide range of halo mass and redshift, some universal properties appear in different dark matter simulations when the halos attain equilibrium \cite{up1,up2,up3, up31, up4,up5,up6}. The density profile of dark matter halos can always be fitted quite well by the famous double power law density profile, i.e, the Navarro-Frenk-White (NFW) model \cite{nfw2,nfw3,nfw4}, which has logarithmic density slope $-1$ close to the center and $-3$ in the outer region. Although there are alternative models (see a brief review \cite{other}), NFW model plays a crucial role in modern dark matter related researches. It finds applications in simulating galaxy formation \cite{rev}, gravitational lensing modeling \cite{app1, app2}, galaxy rotation curves \cite{app3}, dark matter detection experiments \cite{app4}, and so on. In addition, the velocity anisotropy has also been shown to vary as a function of radius \cite{up2,up3,up31}. Another important universal feature in simulations is the so called phase-space density $\rho/\sigma^3$, which seems to follow a power-law with respect to radius \cite{up4,up5}. These universal empirical properties unveil the intrinsic physics of collisionless N-body systems as well as properties of equilibrium dark matter structures in a cosmological setting. However, there is still no universally accepted theoretical model that predicts all of these empirical results \cite{nth0, nth00, nth1, nth2,nshm1,nshm2,nshm3,nshm4}.

Another important aspect is the velocity distributions of dark matter particles. The direct detection experiments of dark matter, which aim to detect the low energy nuclear recoil from rare scattering events between target nuclei and dark matter particles, relies on the knowledge of velocity distributions near the location of solar system. Therefore, it is essential to know the velocity distributions of dark matter halos. The inverse-Eddington method is very efficient to perform this task \cite{ed1}, however, it is only valid under certain limiting circumstances \cite{ed2}. There are also degeneracies in the velocity distributions for a given density profile when the system is not ergodic, because single variable function $\rho(r)$ is not sufficient to determine the velocity distribution which involves multiple variables. An ideal velocity distribution function should also provide the distribution of velocity components. As an example, the standard halo model (SHM) is known to not provide the right description on the velocity distributions \cite{nshm1,nshm2,nshm3,nshm4}, especially the behavior of high velocity tails. Therefore, many velocity distribution models have been proposed, either from the first principle or theories \cite{ vm1,vm2,vma1,vma2,vma3}, guessing from the empirical fitting \cite{vm3,vm4,vm5,vm6,vmt}, inferring from the observational data \cite{vm10,vm11,vm12}, or parameterizing velocity distributions \cite{vm15,vm16,vm17}. However, none of these are fully satisfactory and widely accepted. Furthermore, these velocity distribution models also have degeneracy in the velocity components since they all use the velocity magnitude rather than the velocity components as principle variables.

Instead of working on the density profile and velocity distribution separately, we suggest applying an approach where the density profile (as well as other universal properties) and velocity distributions are consistent with each other. One promising way is to propose a phase space distribution function from which the density profile and velocity distributions could easily be derived and thus automatically be consistent with each other. The density could be obtained through the velocity integral of the phase space distribution function, and the velocity distribution could also be derived by fixing the radius and potential in the phase space distribution function \cite{book}. In previous works, there are action-based phase space distribution that were derived using the action-angel method \cite{aa1,aa2,aa3}. However, the reasoning employed to create the distribution function does not apply to the NFW case. Consequently, there is no action distribution function for NFW density profile, instead they proposed an empirical model for NFW \cite{aa1}. In addition, the action-based models generally do not provide analytical formula for the derived quantities, making their application more challenging. In this work, we propose a new analytical phase space distribution, which could result in a NFW-like density profile, the radial varied anisotropy as well as power law phase-space density (although certain disagreements with simulation data exist). What's more, it can also provide us with the velocity magnitude distribution as well as its components, with natural cut off at the local escape velocity. We will also compare our analytical velocity distributions with simulation data, which has radial and tangential velocity data separately allowing to break the degeneracy of the velocity magnitude distributions.  Our model fit quite well to the radial velocity data and also the low velocity regime of tangential velocity data, giving consistent results on the distributions as well as the fitting parameters within an acceptable range. This provides support that our analytical model may be relevant for understanding the gravitational dynamics of cosmological dark matter structures.

This paper is organized as follows: In Section.\ref{sec2}, we will present our dark matter phase space distribution function and study its predictions on the density profile, anisotropy parameter, as well as the phase space density. Additionally, we will provide the formulas for the velocity magnitude , the radial and tangential velocity components distributions. Next, in Section.\ref{sec3}, we will compare our radial and tangential velocity distributions with simulation data. The velocity data were extracted from different radial bins (shell) of equilibrium dark matter halos, with both radial and tangential velocity data available in each bin. For comparison, we use data from two different simulation schemes, namely the cold collapse and explicit energy exchange. The fitting results and estimated parameters are also presented in this section. Finally, we will conclude and discuss our findings in Section.\ref{sec4}.

\section{The distribution function and its properties}\label{sec2}
Based on the observations of the simulation data introduced in Section.\ref{sec3}, a reasonable distribution function should be spherically symmetric due to the simulated halos have no preferred spatial directions. Therefore, it is convenient to write out the distribution function of dark matter halos in spherical coordinates $(r, \theta, \varphi)$. The distribution function should also be anisotropic because the simulation data of radial and tangential velocity distributions are different. That means we need take into account the angular momentum of dark matter particles. It would be a good choice to put the angular momentum in the exponent because when we integrate the distribution function to obtain the density profile, it will result in a term of $1 + r^{2}$ in the denominator. Then if we wish to obtain an NFW-like profile, we could just multiply it by the radial distance $r$. Meanwhile, we would like the distribution function to have a natural cutoff, and it could be realized by multiplying the binding energy. After considering the above conditions, our distribution function is given by 
\begin{equation}\label{dis}
f(v_r, v_t, r)=f(\epsilon, L, r) \propto \frac{r_0}{r} \times \frac{\epsilon}{\lambda^2} \times \text{Exp}\left(Q/\lambda^2\right)
\end{equation}
where $r_0$ is the parameter to normalize radial distance, and $\lambda$ is the parameter to normalize the binding energy $\epsilon$ and exponent $Q$ which are given by
\begin{align}
\epsilon &= -\frac{v^2}{2} + \phi \\ 
v^2 &= v_r^2 + v_t^2 \\
Q &= \epsilon - L^2/4 r_0^2 \label{Q} \\ 
L & = r v_t
\end{align}
where $\phi$ is the positive potential, $v_r$ and $v_t=\sqrt{v_\theta ^2+v_\varphi ^2}$ are respectively the radial and tangent velocity, and $L$ is the angular momentum of dark matter particles. One could notice that this distribution function does not satisfy the Jeans theorem since there is a factor $r_0/r$, which means that the full form is not only a function of the integrals of motion. This could be seen as the limitation of our model. However, this limitation does not undermine the primary objective of our work. We have proposed a phenomenological or effective model rather than presenting a comprehensive theory. Therefore, it is acceptable for breaching the Jeans theorem as long as it is consistent with the simulation data.

For simplicity, in the following discussion of this section, we will denote $a = r/r_0$ and set $\lambda = 1$. Then the potential, binding energy and $Q$ are all in unit of $\lambda^2$, the velocity is in unit of $\lambda$. 
The density of the above distribution is given by the integral with respect to velocities
\begin{align}\label{den}
\rho (a, \phi) &= 2 \pi \int v_t f(v_r, v_t, a) dv_r dv_t \nonumber\\
&= N_\rho [ \frac{\sqrt{2 \pi } e^{\phi } \left(a^2 (2 \phi -1)+4 \phi -6\right) \text{erf}\left(\sqrt{\phi }\right)}{a \left(a^2+2\right)^2} \nonumber\\
&\quad +  \frac{8 \sqrt{\pi } e^{-\frac{1}{2} \left(a^2 \phi \right)} \text{erfi}\left(\frac{a \sqrt{\phi }}{\sqrt{2}}\right)/a}{a \left(a^2+2\right)^2}+\frac{2 \sqrt{2\phi }}{a \left(a^2+2\right)} ]
\end{align}
where $N_\rho$ is the normalization constant, which is usually the inverse of the halo mass, $erf(x)$ and $erfi(x)$ are called error function and imaginary error function. To obtain $\rho (a)$, we first need to know the dependence of $\phi$ on $a$, which requires solving the spherically symmetric Poisson equation,
\begin{equation}
\frac{d}{da} (a^2 \frac{d \phi}{da}) = -4\pi G a^2\rho(a,\phi)
\end{equation}
however, it has to be solved numerically. Starting from $a = 0$ for some chosen value of $\phi(0)$ and $\frac{d \phi}{da} =0$, we thus can solve for $\phi(a)$.  In practice, we need to apply a softening to the centre point $a=0$ due to the divergent behavior of $1/a$ of the Poisson equation.  If we assume that the central potential is nearly flat within a small radial distance, let's say $a=0.01$. Then, we could solve the Poisson equation by choosing that the initial conditions at $a=0.01$ is the same as the centre point. Thus, the numerical results are not sensitive to the choice of $a=0.01$ as long as the approximation of flat central potential is valid. In the following discussions, we mainly choose to start the numerical solution from $a=0.01$ with different initial potentials, i.e., $\phi(0.01)=5,\, 7,\, 10$ and $\frac{d \phi}{da}|_{a=0.01} =0$. As a comparison, we also show the numerical results of potential when we choose to start with $a=0.001$. The gravitational constant $G$ in the numerical solutions is set to be $4.301 \times 10^{-6}kpc M_{\odot}^{-1}(km/s)^{2}$, and the normalization factor $N_\rho$ is set to $N_\rho =1$.

\subsection{potential, density, anisotropy and pseudo phase pace density}

\begin{figure}
  \includegraphics[scale=0.9]{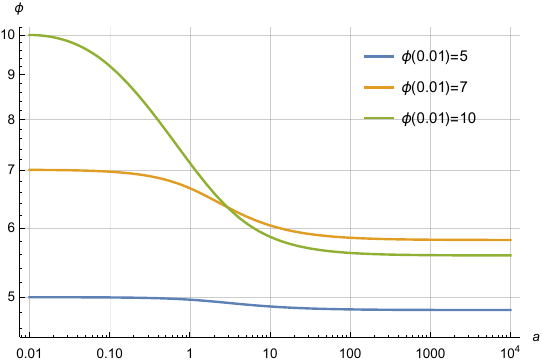}
\caption{The numerical solution of the potentials as a function of radial distance $a$, for three different initial conditions $\phi(0.01)=5,\, 7,\, 10$ and $\frac{d \phi}{da}|_{a=0.01} =0$ at $a=0.01$.}
\label{potential}
\end{figure}

\begin{figure}
  \includegraphics[scale=0.9]{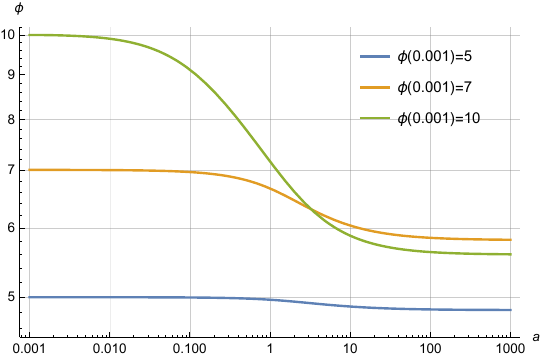}
\caption{The numerical solution of the potentials as a function of radial distance $a$, for three different initial conditions $\phi(0.001)=5,\, 7,\, 10$ and $\frac{d \phi}{da}|_{a=0.001} =0$ at $a=0.001$.}
\label{potential2}
\end{figure}

The numerical results of the potential are shown in Fig.\ref{potential} and Fig.\ref{potential2} for given initial potentials. We can see that all the potentials with different initial conditions are quite smooth when close to the center and at a significant distance from the center, with a drastic transition between them. A larger initial potential, a greater difference between the central and distant regions, and a faster drop in the transition region. By comparing the potentials in Fig.\ref{potential} and Fig.\ref{potential2}, we can see that results are not sensitive to the initial value $a$ in the numerical solutions. The potentials are almost the same regardless of the different initial values of $a$. Therefore, we could safely use one of the numerical results of potential (the case with $a=0.01$) to continue our discussion.

\begin{figure}
\includegraphics[scale=0.9]{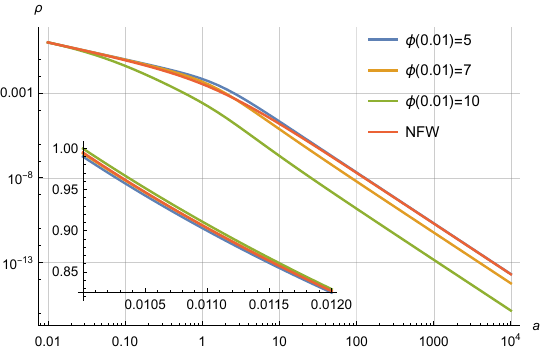}
\caption{The numerical result of dark matter density as a function of radial distance $a$, for three different initial conditions $\phi(0.01)=5,\, 7,\, 10$, and the NFW density profile by setting $r_s=\sqrt{2}\,r_0$. All profiles are normalized by ensuring that they share the same total mass  within their $R_{200}$ radius as $M_{tot}=1$. The concentrations for the curves with $\phi(0.01)=5,\, 7,\, 10$ and NFW are $222.7,\,645.5,\,14866.7$ and $231.9$ respectively.}
\label{density}
\end{figure}

\begin{figure}
\includegraphics[scale=0.9]{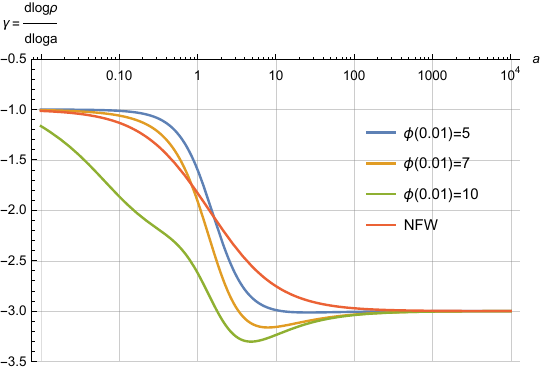}
\caption{The logarithmic slope of the dark matter density $\gamma$ as a function of radial distance $a$, with three different initial conditions $\phi(0.01)=5,\, 7,\, 10$,  and the NFW density slope by setting $r_s=\sqrt{2}\,r_0$.}
\label{slope}
\end{figure}

With the results of potential, we can obtain the density (\ref{den}) corresponding to the distribution function (\ref{dis}). The results are shown in Fig.\ref{density}.  For comparison, we also plot the NFW density profile 
\begin{equation}
\rho_{NFW} \propto \frac{1}{r(r+r_s)^2}
\end{equation}
by setting the characteristic radius of NFW as $r_s=\sqrt{2}r_0$. All the density profiles are normalized by demanding that they have the same total mass $M_{tot}=1$ within their $R_{200}$ radius where the average density within this radius are $200$ times than the mean density of the universe which is set to be $5 \times 10^{-5}$ for simplicity. We can see that they are very similar to the well-known NFW density profile, which has logarithmic density slope $\gamma = {d log \rho}/{d log a}$ approximately $-1$ near the center and $-3$ in the outer region of the halo,  referring to Fig.\ref{slope} for the variation of $\gamma$ with radial distances. However, there are also differences between NFW and our model, especially in the transition region where $\gamma$ is from $-1$ to $-3$. Our model seems more sharper than the NFW density profile around the transition region. The initial potentials will affect the logarithmic slope near the center region. The concentrations $c \equiv R_{200}/R_{-2}$ ($R_{-2}$ is the radius at which the logarithmic slope is $-2$) for the curves in Fig.\ref{density} with $\phi(0.01)=5,\, 7,\, 10$ and NFW are respectively $222.7,\,645.5,\,14866.7$ and $231.9$. It is evident that larger initial potential values result in higher concentrations. To closely resemble the NFW profile, the initial potential should be sufficiently small. The smaller the initial potential, the more similar the profile to the NFW one. It is remarkable that we can obtain the NFW-like density profile from a phase space distribution function.

\begin{figure}
  \includegraphics[scale=0.9]{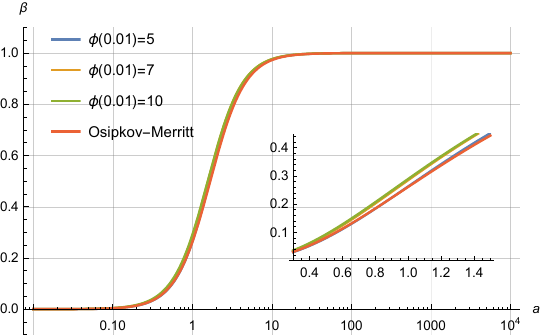}
\caption{The anisotropy parameter $\beta$ as a function of radial distance $a$, with three different initial conditions $\phi(0.01)=5,\, 7,\, 10$,  as well as the $\beta(r)$ of the Osipkov-Merritt model with $r_a^2 =2 \sqrt{2} \, r_0^2$.} 
\label{beta}
\end{figure}

Since our model considers the anisotropic case, it is also important to investigate the anisotropy parameter $\beta$ which is defined as 
\begin{equation}\label{be}
\beta (a) \equiv 1-\frac{\sigma_t^2}{ 2\sigma_r^2}
\end{equation}
where $\sigma_t$ and $\sigma_r$ are respectively the tangential and radial velocity dispersion. They could be compute through 
\begin{align}
\sigma_r^2(a) &= \frac{2 \pi}{\rho(a)} \int v_r^2* v_t f(v_r, v_t, a) dv_r dv_t  \\
\sigma_t^2(a) & = \frac{2 \pi }{\rho(a)} \int v_t^3 f(v_r, v_t, a) dv_r dv_t 
\end{align}
We show $\beta$ as a function of $a$ in Fig.\ref{beta}.
Despite the different initial potentials, we observe that the values of $\beta$ are nearly identical. The initial conditions do not affect the anisotropy. $\beta$ starts at $\beta=0$, i.e., $\sigma_t^2/2=  \sigma_\theta^2 =\sigma_\varphi^2 = \sigma_r^2 $, indicating isotropy around the center. As the radius grows, $\beta$ increase rapidly around the scale radius $a\approx \sqrt{2}$, reaching $\beta=1$ at larger radial distances. This implies $\sigma_t^2 \ll \sigma_r^2 $.  This type of $\beta$ behaves similarly to the commonly-used Osipkov-Merritt model \cite{om1,om2}, which is expressed as
\begin{equation}\label{bop}
\beta(r) = \frac{1}{1+r_a^2/r^2}
\end{equation}
with $r_a$ as the scale parameter in Osipkov-Merritt model. We also illustrate the behavior of $\beta(r)$ when $r_a^2 =2 \sqrt{2}\, r_0^2$ in Fig.{\ref{beta}}. We can see that they are almost the same to our model across a large range of radial distances. This could be attributed to the exponent of our distribution function $Q$ which is similar to the Osipkov-Merritt model.  $\beta(r)$ is only related to the scale parameter $r_a$ and independent of the other parameters of the Osipkov-Merritt model. A similar observation holds for our model, possibly explaining why all the $\beta(a)$ values in our model are almost identical. Eq.(\ref{bop}) could serve as an approximation for $\beta(a)$ in our model, achieved by setting $r_a^2 =2 \sqrt{2} \, r_0^2$. 

\begin{figure}
  \includegraphics[scale=0.9]{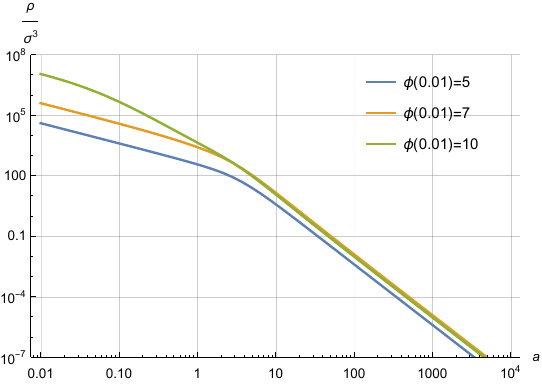}
\caption{The pseudo phase space density as a function of radial distance $a$, with three different initial conditions $\phi(0.01)=5,\, 7,\, 10$.}
\label{psdensity}
\end{figure}

\begin{figure}
  \includegraphics[scale=0.9]{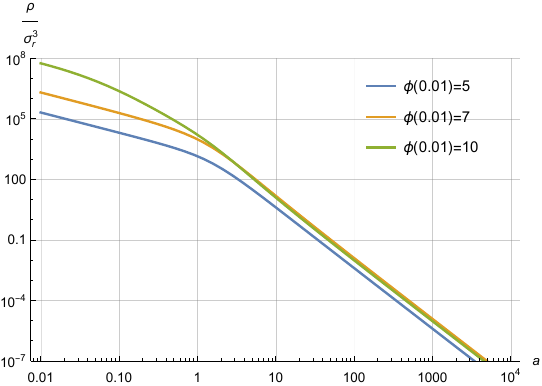}
\caption{The quantity $\rho/\sigma_r^3$ as a function of radial distance $a$, with three different initial conditions $\phi(0.01)=5,\, 7,\, 10$.}
\label{psrdensity}
\end{figure}

We also investigate the pseudo phase space density 
\begin{equation}\label{pe}
\frac{\rho}{\sigma^3}=\frac{\rho}{(\sigma_t^2+\sigma_r^2)^{3/2}}
\end{equation}
which is another important empirical laws. In Fig.\ref{psdensity}, we plot the pseudo phase space density as a function of $a$. We can see that it has a pretty much the same logarithmic density slope as the density, see Fig.\ref{density}.  It has logarithmic slope of $-1$ in the inner region and $-3$ in the outer region. These results disagree with simulation data, which show a simple logarithmic slope $\approx -1.9$ across a large range of orders of magnitude in radius $a$ \cite{up4, up5, add1,add2,add3}. Nevertheless, the different initial potentials only cause a difference in the center region and have no effect on the outer region. For comparison, we also plot the quantity $\rho/\sigma_r^3$ as a function of radius \cite{add1} in Fig.\ref{psrdensity}. The slopes in the outer region are the same as the pseudo phase space density, while there is an overall shift to higher values in the center region.  Unfortunately, these still contradict the simulation data \cite{up5, add1}.

\subsection{velocity distributions}

\begin{figure}
  \includegraphics[scale=0.9]{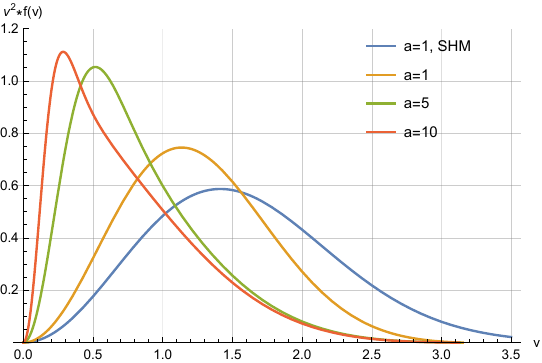}
\caption{The normalized velocity magnitude distribution $v^2f(v)$ as a function of velocity $v$, at three different radial locations $a=1,\,5,\,10$, with initial potential $\phi(0.01)=5$. We also show the Maxwell-Boltzmann distribution which is also known as the standard halo mode (denote as SHM)} at $a=1$, as comparison.
\label{fvelocity}
\end{figure}

\begin{figure}
  \includegraphics[scale=0.58]{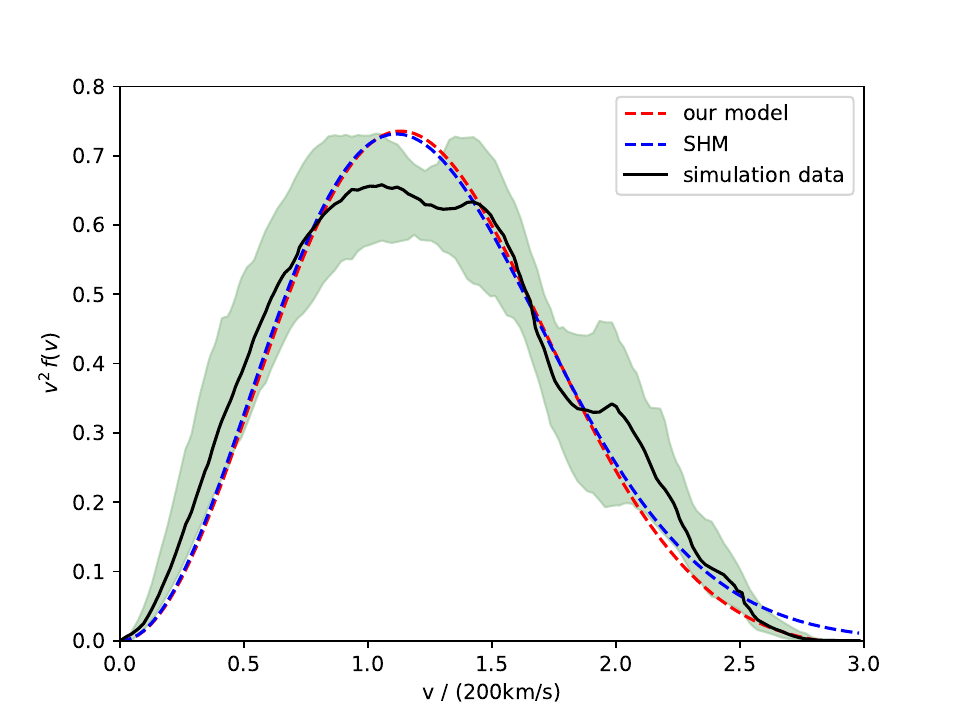}
\caption{The best fit of our model(red dashed) and SHM model(blue dashed) to the velocity distribution data extracted from 2 $kpc$ boxes between 7 and 9 $kpc$ from the center of halo Aq-A-1 in Aquarius Project. The black line is the median value while the green band enclose 95\% of the velocity distributions over all 2 $kpc$ boxes.}
\label{vdata}
\end{figure}

The velocity magnitude distribution $f(v)$ could be obtained by transformation and integrating over the distribution function (\ref{dis}). We get
\begin{equation}\label{fvel}
f(v) = N_v \frac{4 \pi  e^{\phi-\frac{v^2}{2}} \left(2 \phi - v^2\right) \text{DawsonF} \left(\frac{1}{2} \sqrt{a^2 v^2}\right)}{a \sqrt{a^2 v^2}}
\end{equation}
where $N_v$ is the normalization constant, DawsonF$(x)$ is the so-called Dawson integral. We have plotted the velocity magnitude distribution for different radii $a =1,\,5,\,10$ in Fig.\ref{fvelocity}, by choosing the initial potential $\phi(0.01)=5$. As comparison, we also show the SHM velocity distributions which corresponds to the Maxwell-Boltzmann phase space distribution. We can see that our velocity distribution is suppressed more than the SHM model in the high velocity tails. The velocity distributions are more concentrated at lower values as the radial distance becomes larger. We also compare the velocity distribution function (\ref{fvel}) with simulation data from the Aquarius Project \cite{vda1, nshm1}. We fitted the data with our model as well as the SHM model in Fig.\ref{vdata}. The velocity distribution data are taken from many 2 $kpc$ boxes located between 7 and 9 $kpc$ from the center of halo Aq-A-1 in Aquarius Project. In each 2 $kpc$ box, there are about $10^4$ to $10^5$ particles. In Fig.\ref{vdata}, the black line is the median value of velocity distribution measured over all 2 $kpc$ boxes, and the green band encloses 95\% of the measured velocity distributions for each velocity bin, and each velocity bin has a width of 5 $km/s$. We can see that our model almost overlaps the SHM model, with the exception that our model exhibits a relatively larger suppression in the high velocity tails, resulting in a better fit to the simulation data. 

The above velocity magnitude has some extent degeneracy over the velocity components. Therefore, it is crucial to understand the radial and tangential velocity distributions and validate them through simulations.

By integrating over the tangent velocity component in the distribution function (\ref{dis}), this yields the  radial velocity distribution. Similarly, we can integrate over the radial velocity component in the distribution (\ref{dis}) to obtain the tangential velocity distribution. They are respectively given by

\begin{align}\label{rav}
f(v_r) = N_{r} & [ \frac{e^{-\frac{v_r^2-2 \phi }{2 \lambda ^2}} \left(4 \lambda ^2 \left(e^{\frac{\left(a^2+2\right) \left(v_r^2-2 \phi \right)}{4 \lambda ^2}}-1\right)\right)}{a \left(a^2+2\right)^2}\nonumber\\    
& + \frac{e^{-\frac{v_r^2-2 \phi }{2 \lambda ^2}}\left(-v_r^2 \left(a^2+2\right) +2 \phi \left(a^2+2\right) \right)}{a \left(a^2+2\right)^2}  ]  
\end{align}

\begin{align}\label{tanv}
f(v_t) = N_{t} & [ -\frac{e^{-\frac{\left(a^2+2\right) v_t^2}{4 \lambda ^2}} \left(\sqrt{2 \pi } e^{\frac{\phi }{\lambda ^2}} \left(\lambda ^2+v_t^2-2 \phi \right) \text{erf}\left(\frac{\sqrt{\phi -\frac{v_t^2}{2}}}{\lambda }\right)\right)}{2 a \lambda }\nonumber\\    
&+\frac{e^{-\frac{\left(a^2+2\right) v_t^2}{4 \lambda ^2}} \left(2 \lambda  e^{\frac{v_t^2}{2 \lambda ^2}} \sqrt{2 \phi -v_t^2}\right)}{2 a \lambda } ]
\end{align}
where $N_{r}$ and $N_{t}$ represent the normalization constants. To fit the real simulation data in the next section, we have kept $\lambda$ as a free parameter rather than $\lambda=1$ in the above formulas of $f(v_r)$ and $f(v_t)$. It can be verified that the radial and tangential velocity distribution functions approach zero at the local escape velocity $v_r=v_t=v_{escape}=\sqrt{2\phi}$, indicating a cutoff at the local escape velocity.

\section{comparing radial and tangential velocity distributions with simulation data}\label{sec3}

\subsection{the simulation data}

We use the velocity data extracted from the simulation in \cite{sd}. These numerical simulations were described in detail previously \cite{vma1,vma2}. To test our model, we compare the velocity distribution function with two different scheme of simulation performed in \cite{sd}. 

\begin{figure}
  \includegraphics[scale=0.55]{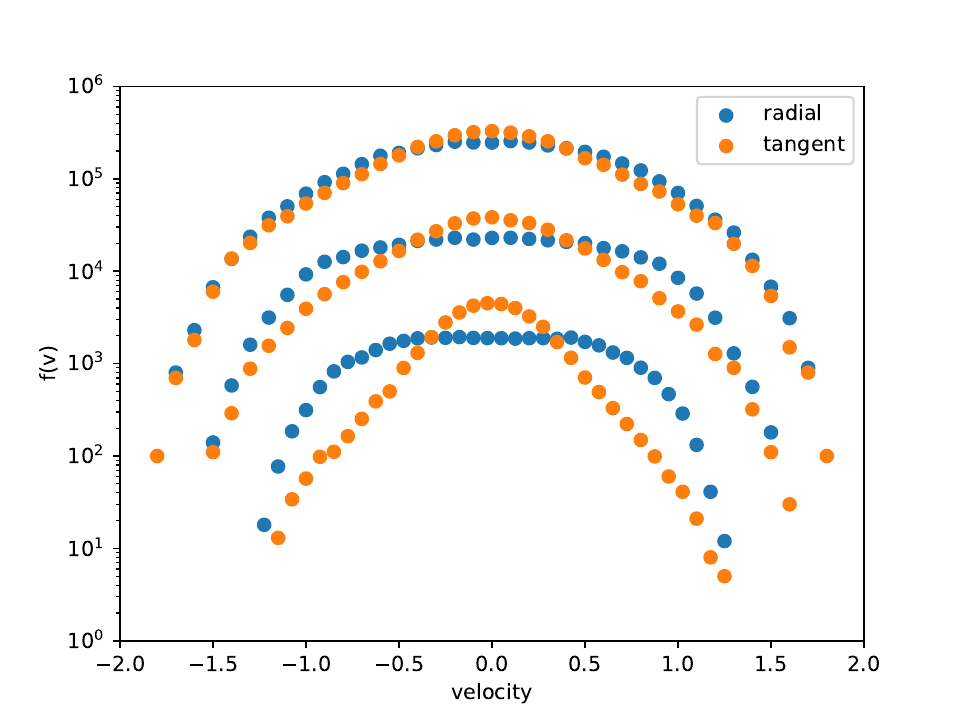}
\caption{The three radial bins data for radial and tangential velocities in the cold collapse simulation. The three radial bins were chosen near the slope of $\gamma_0=-1.6, \, -2.0, \, -2.4$ (From top to bottom), and the $\gamma_0=-1.6, \, -2.0$ data were shifted vertically for easy reading. Each data point represents the number of particles within a velocity bin $\Delta v=0.1$. }
\label{c-data}
\end{figure}

The first is called the cold collapse simulation. This aims to simulate the violent relaxing process that occurs in the early universe when structure collapsed. They create a main halo with a number of compact and condensed substructures. All particles started with zero velocities and evolve under gravity. Once it has attained equilibrium, we divide a halo structure into radial bins (thin spherical shell) and thus we can extract the radial and tangential velocity components of the particles from selected bins. We sample the same number of particles in three radial bins whose radial and the tangential velocity data are shown in Fig.\ref{c-data}. To facilitate the readability of the bin data, some of the bin data have been shifted vertically. The three radial bins were chosen near the slope of $\gamma_0={d log \rho}/{d logr} = -1.6, \, -2.0, \, -2.4$.

\begin{figure}
  \includegraphics[scale=0.55]{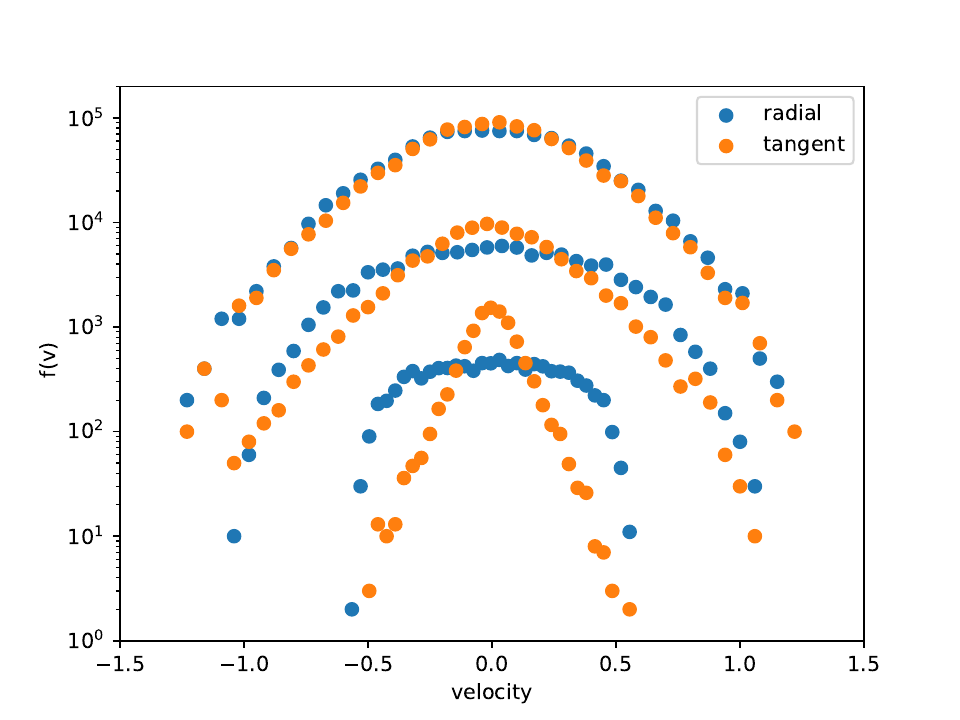}
\caption{The three radial bins data for radial and tangential velocities in the explicit energy exchange simulation. The three radial bins were chosen near the slope of $\gamma_0 = -1.7, \, -2.4\, -3.0$ (From top to bottom), and the $\gamma_0=-1.7, \, -2.4$ data were shifted vertically for easy reading. Each data point represents the number of particles within a velocity bin $\Delta v = 0.07$. }
\label{e-data}
\end{figure}

The second set of numerically simulated data is the so-called explicit energy exchange simulation. Various types of energy exchange take place between collisionless particles, particularly through violent relaxation and dynamical friction. Therefore, the numerical set-up take into account a perturbation in which the spherical symmetry is preserved, but they permit the particles to exchange energy. Each radial bin is designed to preserve energy with instantaneous energy exchange, ensuring that the perturbation itself has no impact on the density or dispersion profiles. Afterward, the system evolves with normal collisionless dynamics. Similarly, we divide the equilibrium structure into radial bins and then extract the radial and the tangential velocity data. We plot the velocity data in Fig.\ref{e-data}. The three radial bins were chosen near the slope of $\gamma_0 = -1.7, \, -2.4\, -3.0$.

\subsection{fitting and parameter estimation for the velocity data}

When fitting our radial and tangential velocity distribution model to the real simulation data, we adopt error-bars on these data. We assume that the horizontal (velocity) error for each data point corresponds to the width of velocity bin $\Delta v$, while the vertical (frequency) errors are taken as the square root of the counts in each velocity bin. This assumption is based on considering the particles in each velocity bin follow a Poisson distribution. In total, we have four parameters to fit for radial or tangential velocity distribution data: the normalization factor $N_r (N_t)$, the radial distance $a$, the potential $\phi(a)$, the parameter $\lambda$.

From an ideal perspective, if our model is the truth, when fitting the radial and tangential data independently in each bin, there should be two constraints for the fitting parameters of the data: first, since the three radial bins data are taken from one halo structure, the value of $\lambda$ for the three radial bins should be the same; second, for each radial bin, the fitting parameters ($a$, $\phi$) for the radial and tangential data should also be the same.  Instead of fitting the radial and tangential data independently, we choose to jointly fit the radial and tangential data for each bin by using the same values of $a$ and $\phi(a)$ in the fitting algorithm, i.e. we define the total logarithmic likelihood function as the sum of the radial and tangential parts
\begin{align}
&ln \mathcal{L}(\text{rad and tan data}|{N_r, N_t, a, \phi(a), \lambda})\\
&=  ln \mathcal{L}_1(\text{rad data}|{N_r, a, \phi(a), \lambda})+ ln \mathcal{L}_2(\text{tan data}|{N_t, a, \phi(a), \lambda})\nonumber
\end{align}
where \texttt{ran} and \texttt{tan} data represent the radial and tangential velocity data in each radial bin. With this joint fitting, the second constraint could be easily implemented. Then, we employ Bayesian inference with flat priors and MCMC sampling method to explore the posterior distributions of parameters given the velocity data from each radial bin. We use 300 random walkers, and each walker takes 5000 steps.  After excluding 10\% of the samples in the burning phase, we finally obtain the samples that can be used to analyze our model.

The fitting parameters for each bin in the collapse simulation data are presented in Table.\ref{c-para}. We show the median values as well as the upper and lower bound that enclose 68\% of the MCMC samples. The normalization factors increase with $\gamma_0$, showing a positive correlation. The median values of radial distance $a$ and potential $\phi(a)$ are reasonable, as they exhibit growth and decrease respectively with increasing $\gamma_0$. Additionally, for comparison with the fitting potential, we also show the local escape kinetic energy $v_{max}^2/2$ for each bin, where $v_{max}$ represents the maximum velocity ($\approx$ local escape velocity) of each bin. Ideally, the fitting potential should be equal to local escape kinetic energy, represented by $\phi \approx v_{max}^2/2$. We could see from Table.\ref{c-para} that the fitting potential is approximately equal to the local escape energy. This supports the self-consistency of the fitting $\phi(a)$ in different radial bins. It is noteworthy that our model appears to violate the first constraint mentioned above, as the fitting $\lambda$ values are not uniform. However, in practice, due to the size of radial bins and the limitation of particle numbers in the simulation, some flexibility in adhering to these constraints is expected.

\begin{table}
\setlength{\tabcolsep}{3mm}
\renewcommand{\arraystretch}{1.5}
\begin{tabular}{cllllllll}
\hline \hline
\multirow{2}{*}{} & \multicolumn{2}{c}{$\gamma_0=-1.6$} & \multicolumn{2}{c}{$\gamma_0=-2.0$} & \multicolumn{2}{c}{$\gamma_0=-2.4$} \\ \hline
$N_r$      &    \multicolumn{2}{c}{$185.114\,^{+34.140}_{-27.785}$}  &    \multicolumn{2}{c}{$122091\,^{+67844}_{-50666}$}  &    \multicolumn{2}{c}{$1064057\,^{+678118}_{-448879}$} \\ 
$N_t$    &    \multicolumn{2}{c}{$38.782\,^{+7.105}_{-5.800}$}  &    \multicolumn{2}{c}{$21814\,^{+12125}_{-9034}$}  &    \multicolumn{2}{c}{$115227\,^{+73459}_{-48646}$}\\
$a $    &    \multicolumn{2}{c}{$1.033\,^{+0.033}_{-0.030}$}  &    \multicolumn{2}{c}{$4.601\,^{+0.635}_{-0.643}$}  &    \multicolumn{2}{c}{$9.175\,^{+1.480}_{-1.361}$}\\ 
$\phi(a)$   &    \multicolumn{2}{c}{$1.684\,^{+0.024}_{-0.024}$}  &    \multicolumn{2}{c}{$1.095\,^{+0.014}_{-0.010}$}  &   \multicolumn{2}{c}{$0.724\,^{+0.007}_{-0.006}$} \\ 
$v_{max}^2/2$   &    \multicolumn{2}{c}{$1.805$}  &    \multicolumn{2}{c}{$1.280$}  &   \multicolumn{2}{c}{$0.781$} \\ 
$\lambda$   &    \multicolumn{2}{c}{$0.770\,^{+0.010}_{-0.009}$}  &    \multicolumn{2}{c}{$1.953\,^{+0.259}_{-0.259}$}  &    \multicolumn{2}{c}{$1.996\,^{+0.318}_{-0.292}$} \\  \hline\hline
\end{tabular}
\caption{The best estimated parameters for the three radial bins in the collapse simulation. We show the median values with upper and lower bounds enclosing 68\% of the MCMC samples. Additionally, we display the local escape energy $v_{max}^2/2$ for each bin as reference to compare with the potential. } 
\label{c-para}
\end{table}

\begin{figure}
  \includegraphics[scale=0.46]{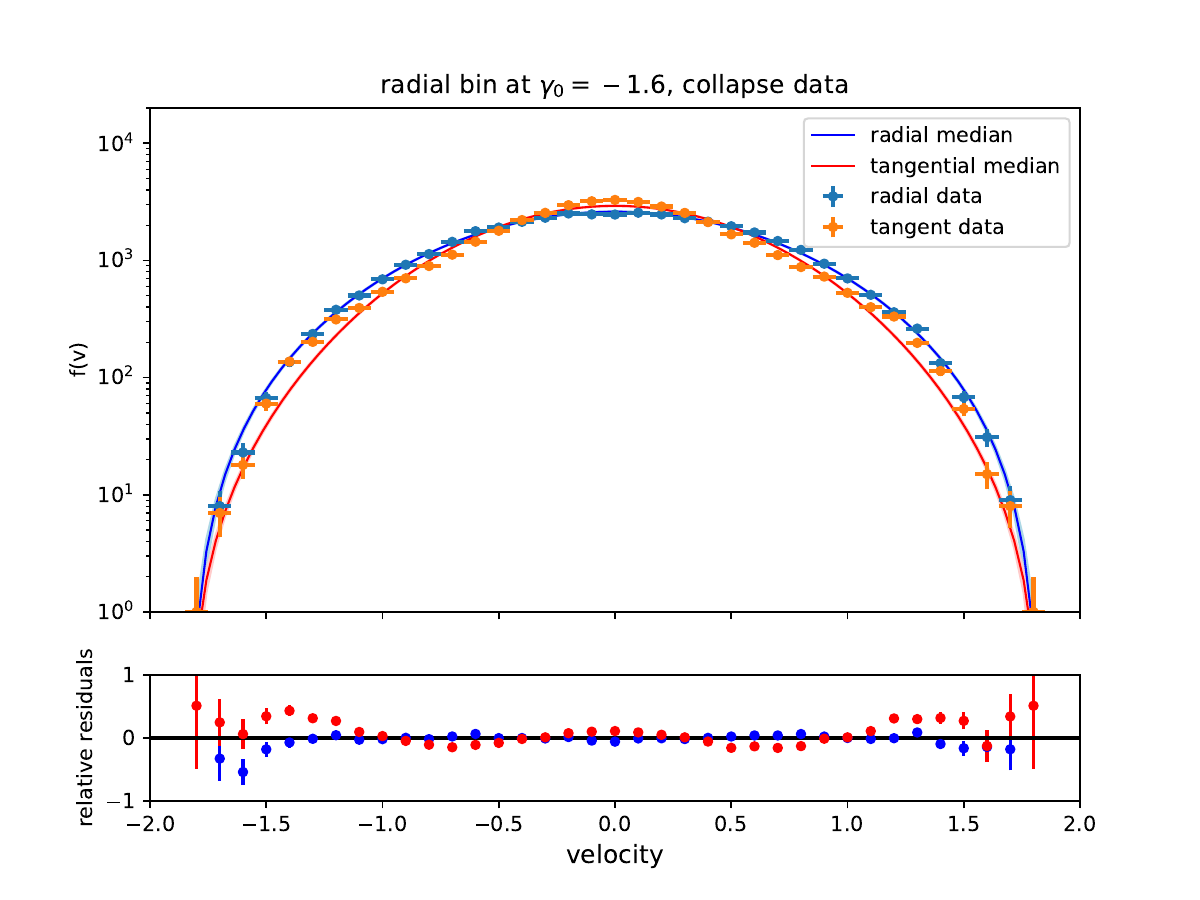}
  \includegraphics[scale=0.46]{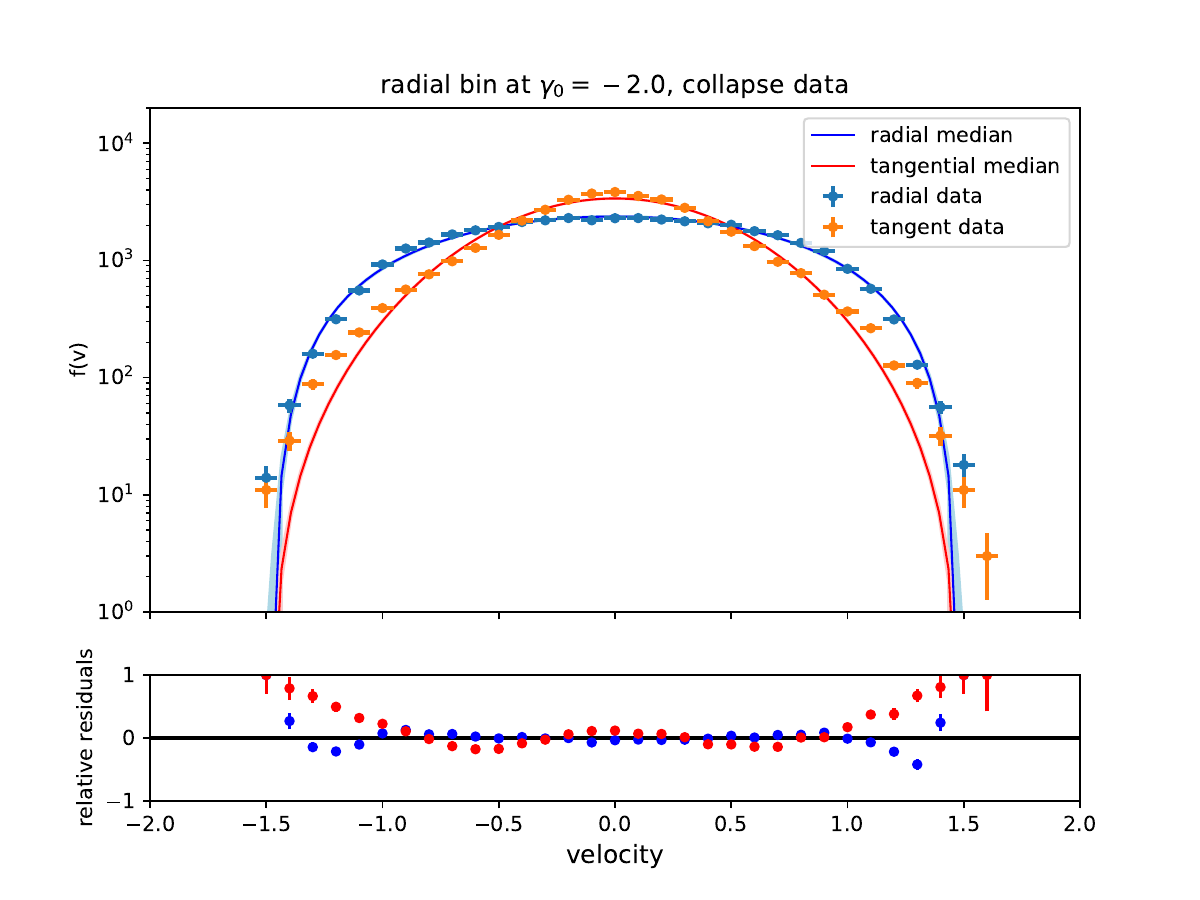}
  \includegraphics[scale=0.46]{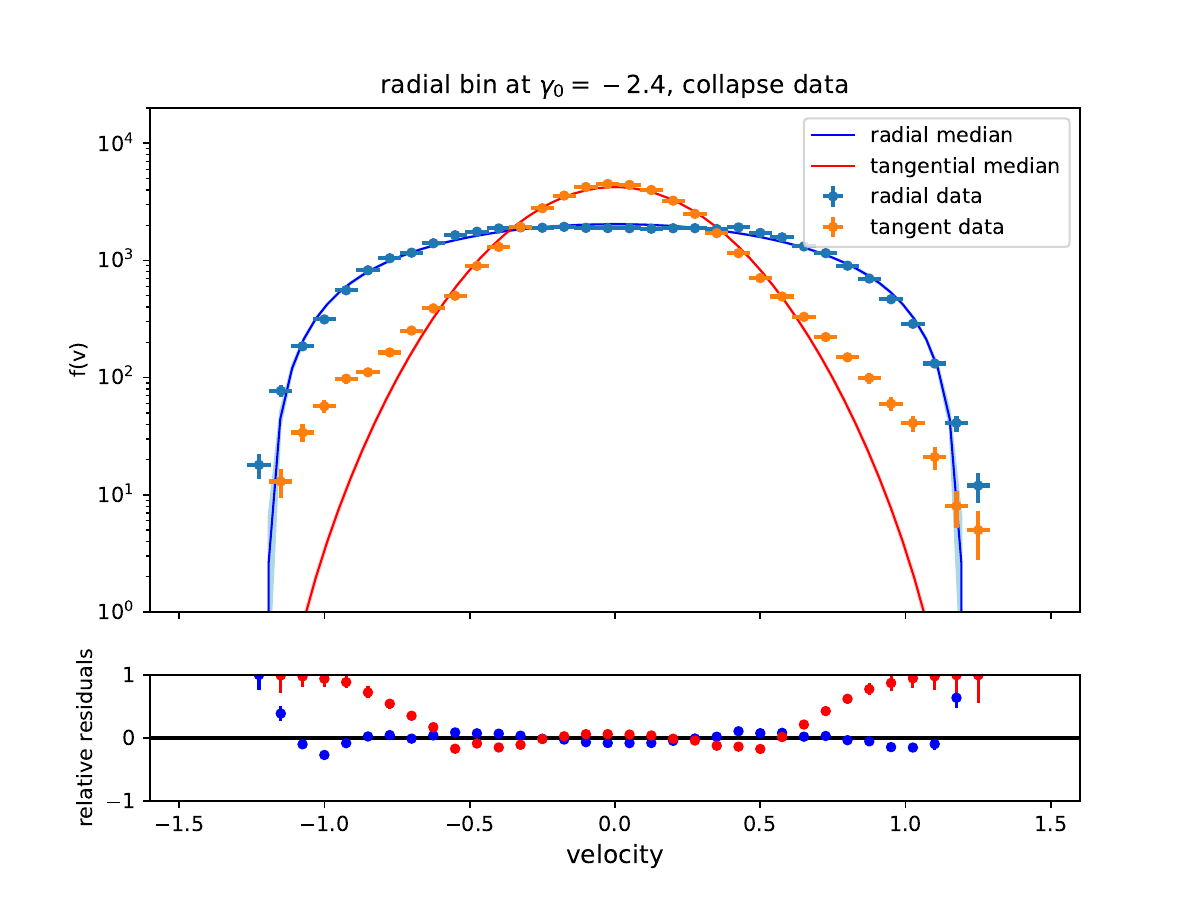}
\caption{The radial and tangential velocity fitting results to the three radial bins in the cold collapse simulation. From top to bottom are respectively $\gamma_0= -1.6, \, -2.0, \, -2.4$. The radial data and the median radial distributions are colored blue, while the tangential results are colored red. The 1-sigma band of radial distributions are colored shallow blue, and tangential bands are colored shallow red. The relative residuals are also given for the radial (colored blue) and tangential data (colored red).}
\label{c-fits}
\end{figure}

To be more explicit, we present the fitting results to the three radial bins of collapse simulation in Fig.{\ref{c-fits}}. From top to bottom, we present the data and fits for the bins with $\gamma_0= -1.6, \, -2.0, \, -2.4$. In each plot, we show both the radial and tangential velocity data along with the corresponding error-bars, as mentioned above. The solid blue curve represents the radial velocity distribution, while the solid red curve represents the tangential velocity distribution, both predicted by the median values of the MCMC samples. Additionally, we include the 1-sigma band of the velocity distribution predicted by the MCMC samples. The band is colored in shallow blue for the radial fit and shallow red for the tangential fit, though it may be relatively small for some bins. The relative residuals, defined as the difference between the data and the solid curves, are calculated as
\begin{equation}
\frac{\text{data} - \text{solid curve}}{\text{data}}
\end{equation}
are also shown in the bottom of each subplot in Fig.{\ref{c-fits}}. The blue and red dots respectively represent the radial and tangential relative residuals. We can see from Fig.{\ref{c-fits}}, our model gives a good fit to all three bins especially for the radial data, despite some residuals remaining. The radial fits of our model are much better than the tangential fits. Our model only performs well in the low velocity regime of tangential data and underestimate their high velocity tails. The relative residuals increase as $\gamma_0$ and velocity increase. Nevertheless, as expected, the radial and tangential velocity converge to each other at higher velocities. The smaller the value of $\gamma_0$, the higher  the similarity of radial and tangential velocity distributions.


\begin{table}
\setlength{\tabcolsep}{3.3mm}
\renewcommand{\arraystretch}{1.5}
\begin{tabular}{cllllllll}
\hline \hline
\multirow{2}{*}{} & \multicolumn{2}{c}{$\gamma_0=-1.7$} & \multicolumn{2}{c}{$\gamma_0=-2.4$} & \multicolumn{2}{c}{$\gamma_0=-3.0$}  \\ \hline
$N_r$      &    \multicolumn{2}{c}{$2.776\,^{+6.585}_{-1.559}$}  &    \multicolumn{2}{c}{$3107\,^{+762}_{-913}$}  &    \multicolumn{2}{c}{$83211678\,^{+393502366}_{-71717150}$} \\ 
$N_t$    &    \multicolumn{2}{c}{$0.363\,^{+0.855}_{-0.204}$}  &    \multicolumn{2}{c}{$342.9\,^{+84.6}_{-100.2}$}  &    \multicolumn{2}{c}{$3249348\,^{+15434273}_{-2798681}$}\\
$a $    &    \multicolumn{2}{c} {$0.737\,^{+0.070}_{-0.057}$}  &    \multicolumn{2}{c}{$2.370\,^{+0.112}_{-0.132}$}  &     \multicolumn{2}{c} {$38.039\,^{+29.741}_{-18.075}$} \\ 
$\phi(a)$   &  \multicolumn{2}{c} {$0.936\,^{+0.078}_{-0.104}$}  &    \multicolumn{2}{c}{$0.539\,^{+0.017}_{-0.007}$}   &    \multicolumn{2}{c}{$0.164\,^{+0.001}_{-0.001}$}  \\ 
$v_{max}^2/2$   &  \multicolumn{2}{c} {$0.832$}  &    \multicolumn{2}{c}{$0.562$}   &    \multicolumn{2}{c}{$0.174$}  \\ 
$\lambda$   &    \multicolumn{2}{c} {$0.404\,^{+0.013}_{-0.007}$}  &  \multicolumn{2}{c} {$0.631\,^{+0.023}_{-0.029}$}  &    \multicolumn{2}{c} {$2.875\,^{+2.259}_{-1.361}$} \\  \hline\hline
\end{tabular}
\caption{The best estimate parameters for the three radial bins in the explicit energy exchange simulation. We show the median values  along with upper and lower bounds that enclose 68\% of the MCMC samples. In addition, we also show the local escape energy $v_{max}^2/2$ for each bin as reference for comparison with the potential.} 
\label{e-para}
\end{table}

\begin{figure}
  \includegraphics[scale=0.46]{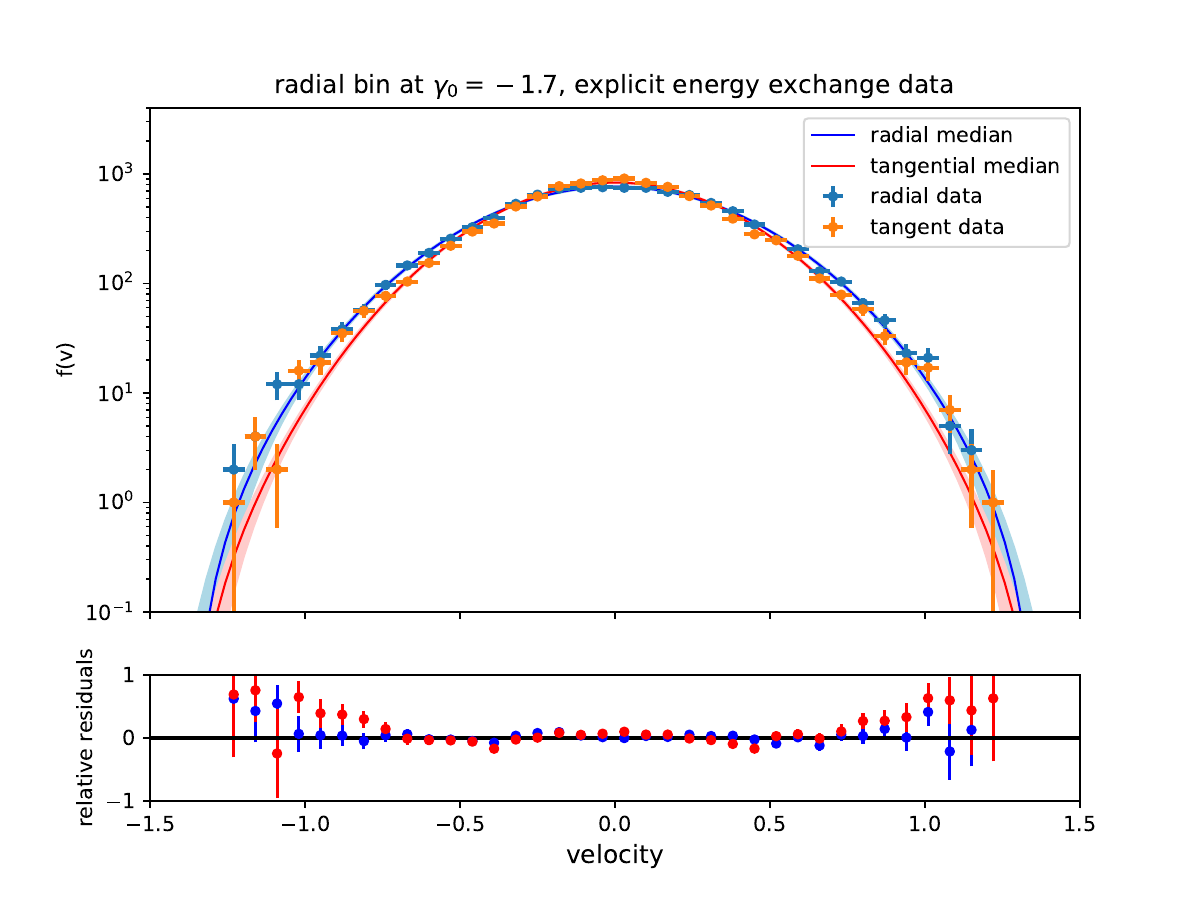}
  \includegraphics[scale=0.46]{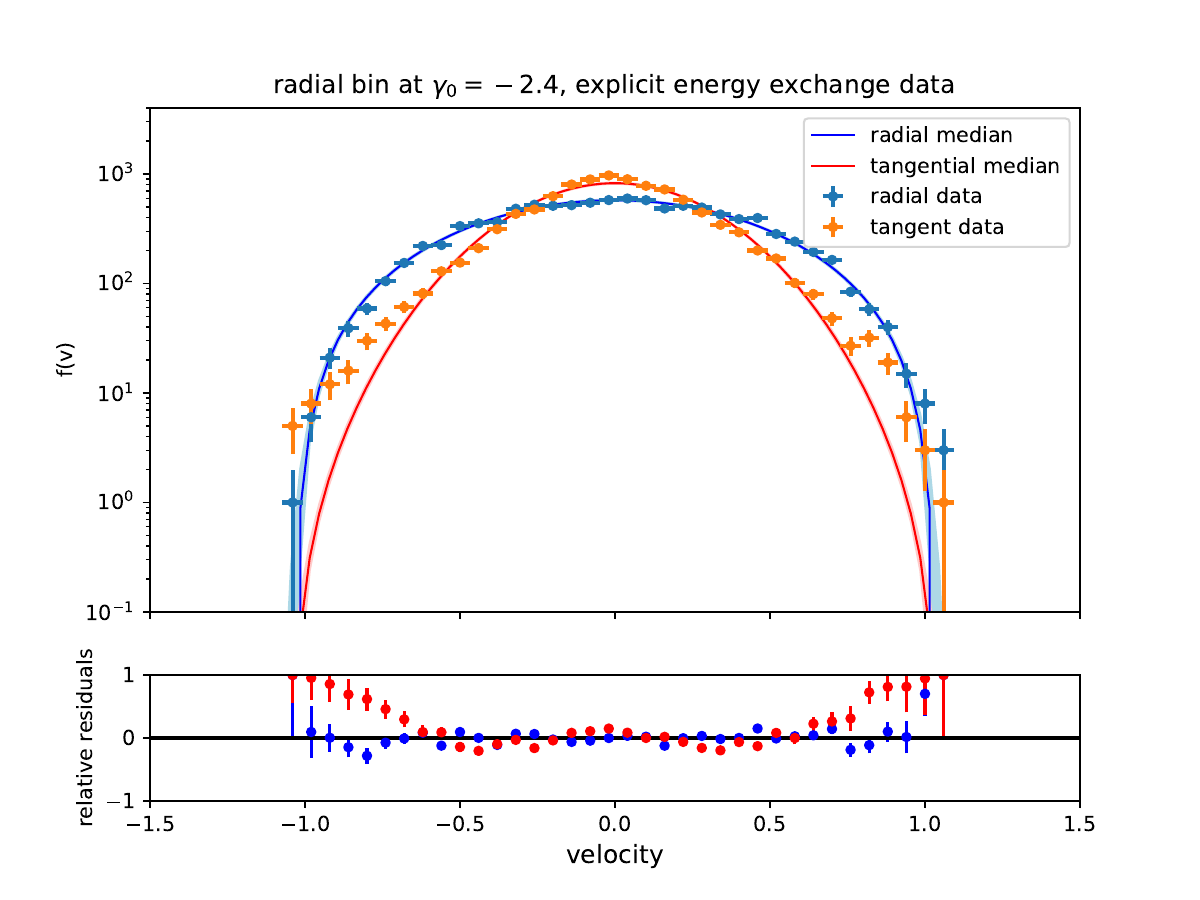}
  \includegraphics[scale=0.46]{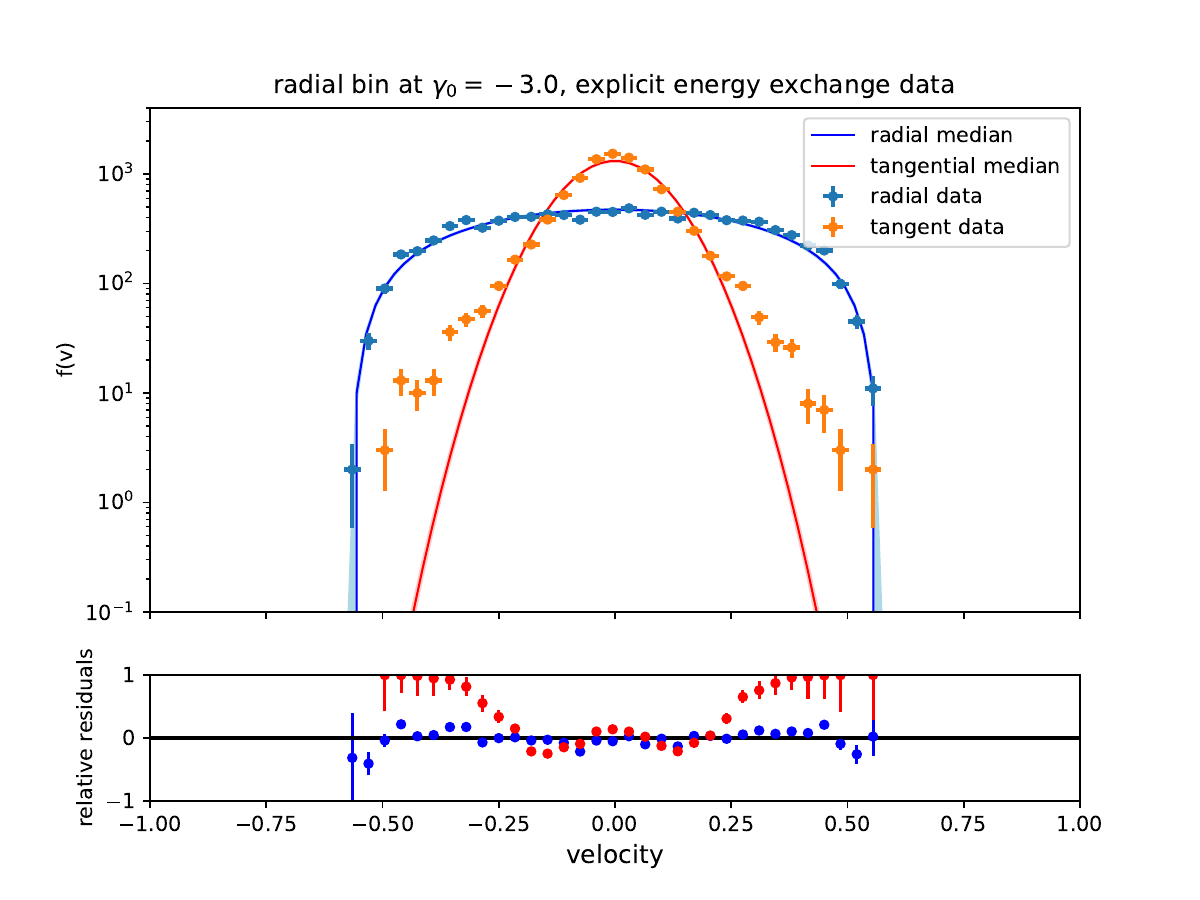}
\caption{The radial and tangential velocity fitting results to the three radial bins in the explicit energy exchange simulation. From top to bottom are respectively $\gamma_0 = -1.7, \, -2.4\, -3.0$. The radial data and the median radial distributions are colored in blue, while the tangential results are colored in red. The 1-sigma band of radial distributions are colored in shallow blue, and tangential bands are colored in shallow red. The relative residuals are also given for the radial (colored in blue) and tangential data(colored in red).}
\label{e-fits}
\end{figure}

We have also shown the fitting parameters of the three radial bins for the explicit energy exchange simulation data in Table.\ref{e-para}. Similar to the situation with the collapse simulation data, we also show the upper and lower bounds that encompass 68\% of the MCMC samples, together with the median values. As $\gamma_0$ rises, so do the normalization factors. The median of radial distance $a$ and potential $\phi(a)$ are following a good trend along with $\gamma_0$. The fitting $\phi(a)$ in different radial bins are also self-consistent since $\phi \approx v_{max}^2/2$ in each bin. The fitted bands of the radial bin at $\gamma_0=-3$ is quite large compare to the other two radial bins. In addition, because fitting $\lambda$ vary among different bins, we can also observe that our model appears to break the first constraint, similarly to the case in the collapse simulation.

We also plot the fitting results for the explicit energy exchange simulation data in Fig.{\ref{e-fits}}. From top to bottom are respectively the $\gamma_0= -1.7, \, -2.4, \, -3.0$ bins data and fits. As in the collapse simulation case, we also show the tangential and radial velocity data along with the previously mentioned error bars in each plot. The radial and tangential velocity distributions predicted by the median value of the fitting parameters are also shown by the solid blue and red curves. We present the 1-sigma band and relative residuals with the same setup in the collapse simulation case.
We can also see from Fig.{\ref{e-fits}} that the situation is almost the same to the collapse simulation case. The patterns are repeating. Our model provides a fair fit to all three bins, particularly for the radial data, while it fails again on predicting the high velocity tails of the tangential data. As $\gamma_0$ and velocity increase, the relative residuals also increase.

Although there are slight differences, the similarity of the data and fits in both cold collapse and explicit energy exchange simulation reveals an universal dark matter velocity distributions. The basic trends and structures of dark matter velocity distributions were captured by our model. The data suggests that our model could be relevant to describe the distribution of dark matter particles,  especially for the radial velocity distributions. In principle, we could use the fitted $a$ and the normalized potential $\phi(a)/\lambda^2$ of three radial bins in each simulation to reproduce the density profiles of the original simulated halo. However, the discrepancy of fitting $\lambda$ to three bins add large uncertainties on the normalized potential $\phi(a)/\lambda^2$. So, we will leave this as a problem that needs to be addressed in the future.

\section{conclusion and discussion}\label{sec4}

In this paper, we have suggested and analysed a phenomenological model of spherically symmetric anisotropy dark matter phase space distribution function as presented in Eq.(\ref{dis}). We first introduced our distribution function and demonstrated their predictions on integrated quantities, which are typically analysed in numerical simulations, namely the density profile (\ref{den}), the anistropy (\ref{be}) and phase space density (\ref{pe}). In addition, we also give the velocity magnitude (\ref{fvelocity}), the radial and tangential velocity distributions in analytical forms (\ref{rav}) (\ref{tanv}). To assess the utility of our velocity distributions, we have respectively compared with radial and tangential velocity data which were extracted from different radial bins of equilibrium dark matter halos. We verify our results in two different simulation schemes, the cold collapse and explicit energy exchange which effectively cover a wide range of possible equilibrium processes. The fits show good performance and the results are promising, especially for the radial data (across the whole velocity range) and the lower velocity regime of tangential data. There is agreement in estimating the relevant parameters of the radial and tangential velocity data.

Despite the advantages of providing good fits to the velocity distributions, there are also some potentially important limitations to our model that we wish to address. Our model faces the first challenge in predicting the pseudo phase space density, which does not follow a single, unbroken power law. The second limitation arises when comparing with the data, as significant deviations are observed, especially in the high-velocity tails of tangential data. Additionally, the predicted parameter $\lambda$ differs among three bins of the same halo. These observations could indicate that our model is missing some additional features of dark matter halos. However, it is essential to verify this issue using another independent simulation program since the error could originate from the data rather than the model. If these limitations have been confirmed by other simulations, we should explore possible ways to improve our model. Our model only uses the linear order of binding energy $\epsilon$ and exponent $Q$, which represents the simplest case. One could consider using $\epsilon^n$ or $Q^n$ in the exponent of (\ref{dis}) with $n\ge 0$. Introducing non-linearity could potentially compensate for these limitations. Furthermore, the $r_0/r$ factor in our phase space distribution (\ref{dis}) could also be a simplification of some complicated function involving integrated quantities such as $\epsilon$ and $Q$. If $r_0/r$ could be replaced by functions of integrated quantities, the model would be much plausible, indicating its adherence to the Jeans theorem and transcending a mere phenomenological model.

\section*{acknowledgments}
Zhen Li would like to acknowledge the DARK cosmology center at the Niels Bohr Institute for supporting this study. Zhen Li is also financially supported by the China Scholarship Council. Zhen Li would also like to express gratitude to HongSheng Zhao for the helpful discussions that have significantly improved this work.
\\

\section*{Data Availability}
The data used in this study are available upon request. Researchers interested in accessing the data can contact the corresponding author (CA) for more details.

\bibliographystyle{mnras}
\bibliography{fh}

\bsp	
\label{lastpage}
\end{document}